\documentclass[reprint,superscriptaddress,aps,pra]{revtex4-1}

\usepackage[T1]{fontenc}
\usepackage{xcolor}
\usepackage{etoolbox} 
\usepackage{graphicx}
\usepackage{amsmath}
\usepackage{amssymb}
\usepackage{bm}
\usepackage[colorlinks=true,
            linkcolor=blue,
            citecolor=blue,
            urlcolor=blue]{hyperref}

\begin{document}

\title{Local conditional information in anomalous diffusion beyond second-order statistics: A masked generative measurement}

\author{Yu Zhang}
\affiliation{School of Physics and Electronics, Hunan University, Changsha 410082, China}
\author{Zihan Huang}
 \email{huangzih@hnu.edu.cn}
\affiliation{School of Physics and Electronics, Hunan University, Changsha 410082, China}


\begin{abstract}
The relation between the increments of a stochastic trajectory is commonly described by second-order statistics such as the mean squared displacement, but for non-Gaussian processes it extends beyond correlations. We define local conditional information as a general measure of this relation: the expected score improvement that a trajectory's neighborhood provides for a hidden stretch of increments. It decomposes into a second-order part given by kriging (Gaussian-process regression) and a remainder beyond second order. To estimate this quantity without assuming a generating mechanism, we train conditional generative models by masked learning, with the sole objective of reconstructing hidden stretches from their trajectory neighbors. Under a strictly proper scoring rule, the resulting estimate is a lower bound on the true information, so any excess over kriging's gain establishes the presence of information beyond second order. We evaluate this measurement on the five canonical mechanisms from the Anomalous Diffusion challenge and two additional Brownian-yet-non-Gaussian processes. The estimate reproduces kriging's gain for the Gaussian processes and recovers 75\%--99\% of the information beyond second order that follows from the non-Gaussian mechanisms. Brownian motion and the two Brownian-yet-non-Gaussian processes have identical second-order statistics, yet a single neighbor on either side of each hidden stretch raises the accuracy of assigning single trajectories to their process from 62\% to 94\%. On experimental trajectories of receptors in the membrane of living cells, for which no mechanism is assumed, the estimated information lies almost entirely beyond second order. A general theory reveals the origin of the local conditional information beyond second order: a state of the particle or of its surroundings that persists in time, such as a diffusivity or a velocity. The learned conditional densities also enable unsupervised trajectory segmentation without changepoint labels: for example, per-increment mechanism label recall for the continuous-time random walk rises from 2.5\% with second-order evidence to 99.8\%. The same procedure segments experimental trajectories of quantum dots transported by kinesin motors. Together, these results establish local conditional information as a measurable property of stochastic processes that reaches beyond second-order statistics, and point to masked learning as a general paradigm for discovery in physics.
\end{abstract}

\maketitle

\section{Introduction}\label{introduction}

Anomalous diffusion~\cite{MetzlerKlafter2000,HoflingFranosch2013,Metzler2014PCCP}, characterized by deviations from classical Brownian motion, is widespread in diverse systems, such as living cells~\cite{GoldingCox2006,Sabri2020}, complex fluids~\cite{Wong2004,LevinBelRoichman2021}, and financial markets~\cite{Plerou2000}. Its description, like that of any diffusion process, rests on the statistics of the increments and, in particular, on the relation between increments at different times. This relation is commonly described by second-order statistics, including the covariance function, the mean squared displacement (MSD) and its associated scaling exponent $\alpha$, the velocity autocorrelation function (VACF), and the power spectral density (PSD)~\cite{HoflingFranosch2013,Weber2010,Krapf2019PRX,Sposini2020}. For Gaussian processes like fractional Brownian motion (FBM)~\cite{MandelbrotVanNess1968} and scaled Brownian motion (SBM)~\cite{LimMuniandy2002,Jeon2014SBM}, the mean and covariance functions fix all finite-dimensional distributions, so second-order statistics are complete. For non-Gaussian processes~\cite{BelNemenman2009,Wang2009,Akimoto2011,ChubynskySlater2014,Chechkin2017,LanoiseleeGrebenkov2018,BarkaiBurov2020}, however, the relation between increments extends beyond their correlations. A waiting period~\cite{MontrollWeiss1965,ScherMontroll1975}, a patch of constant diffusivity~\cite{Massignan2014}, or a persistent flight~\cite{ShlesingerWestKlafter1987,KlafterBlumenShlesinger1987} is a relation between a stretch of trajectory and the stretches around it. The neighbors tell whether the particle is still waiting, how fast it currently diffuses, or where it is heading. Conventional diagnostics of non-Gaussian behavior, such as the non-Gaussian parameter, the $p$-variation, and the codifference~\cite{Rahman1964,Magdziarz2009,Wylomanska2015}, compress the trajectory into a summary in which this temporal relation is largely lost. Model-based inference~\cite{Masson2009,Persson2013,Thapa2018} and supervised machine learning~\cite{Granik2019,MunozGil2020,Argun2021,WADNet2021,GroupReview2025,GroupWavelet2026} recover this relation by identifying the mechanism, but they rely on candidate mechanisms~\cite{Presse2015FFPE,Tsekouras2015FCS} and labeled data, respectively. What is still missing is a general description of this relation, obtained from trajectories alone.

Such a description is suggested by the examples above: how much the neighbors of a stretch tell about it. We call this amount the {\it local conditional information}. Unlike the predictive information, which measures how much the past of a time series tells about its future~\cite{BialekNemenmanTishby2001}, it concerns a finite stretch and its neighbors on both sides. It quantifies the relation between increments at every order and for any process, and thus extends the description of stochastic processes beyond second-order statistics. For a Gaussian process, second-order statistics already provide it: the answer is given by kriging~\cite{Krige1951mine,Matheron1963}, the optimal interpolator for Gaussian processes, known in machine learning as Gaussian-process regression~\cite{RasmussenWilliams2006}. For a stretch of increments hidden in a trajectory, kriging gives its distribution conditioned on the increments on either side, which is again Gaussian, with a mean linear in the neighbors and a covariance independent of them. It requires only the mean and covariance function of the process. For a non-Gaussian process, however, kriging recovers only the second-order part of the information. This limitation is most apparent for two processes with identical second-order statistics. Kriging predicts a hidden stretch in the same way for both, even when the information that the neighbors provide about it differs completely between them. That difference lies beyond second-order statistics, in the true conditional distributions of the hidden stretch. Measuring this part of the information means solving the same interpolation problem without the Gaussian assumption, with the conditional distribution learned from the trajectories themselves.

Such learning from the data alone, without labels, has a striking precedent in language modeling. Language models are pretrained by masking part of a text and predicting it from the rest~\cite{Devlin2019,Manning2020PNAS}, and this signal alone has proved sufficient to learn the structure of language. A trajectory can be treated in the same way. Its increments carry no label, and the non-Gaussian features described above are relations between a stretch and the stretches around it, which is what a model must learn in order to fill a gap. Masking a stretch of increments and reconstructing it from the increments on either side turns the interpolation problem of the previous paragraph into a training objective. The hidden stretch itself is the target, so neither candidate mechanisms nor labeled data are required. Therefore, masked learning is a direct route to the problem posed above: it needs the trajectories of a process and nothing else, and what it learns is the conditional distribution that the measurement calls for.

In this work, we introduce the local conditional information and measure it by masked learning. Kriging's gain is the second-order part of this information. With a strictly proper scoring rule~\cite{GneitingRaftery2007}, imperfect training can only lower the estimate; whatever the model recovers in excess of kriging's gain is therefore information that second-order statistics cannot supply, for any generating mechanism. On the five canonical mechanisms of the Anomalous Diffusion (AnDi) challenge~\cite{AnDi2021}, the estimate reproduces kriging's gain for the Gaussian processes, whereas the non-Gaussian mechanisms carry information beyond second order. On two Brownian-yet-non-Gaussian processes~\cite{ChubynskySlater2014,Chechkin2017,MiyaguchiAkimotoYamamoto2016}, which share the covariance of Brownian motion, the recovered information lies entirely beyond second order and allows single trajectories to be assigned to the process that generated them. On experimental trajectories of membrane receptors in living cells~\cite{Lanoiselee2021}, with no mechanism assumed, the recovered information also lies almost entirely beyond second order. A general theory reveals the origin of this information: a persistent state of either the particle or its surroundings, such as a diffusivity or a velocity. Conditional densities learned in the same way also locate mechanism changes along a multi-mechanism trajectory without changepoint labels, and experimental trajectories of quantum dots transported by kinesin motors~\cite{Katrukha2017} are segmented by the same procedure. Local conditional information is thus a measurable property of a diffusion process, and masked learning provides its measurement from trajectories alone.

The rest of this paper is organized as follows. Sec.~\ref{sec:framework} defines the local conditional information, separates its second-order part, and presents its measurement by masked learning together with the lower-bound property. Sec.~\ref{sec:results} reports the measurement for Gaussian processes, for non-Gaussian processes, and for processes with identical second-order statistics, and derives the theoretical origin of the information. It closes with the measurement on experimental trajectories. Sec.~\ref{sec:segmentation} uses the learned conditional densities to segment simulated and experimental trajectories without changepoint labels. Sec.~\ref{conclusion} concludes and outlines further directions. The Supplemental Material (SM)~\cite{SupplementalMaterial} (see also references~\cite{Ba2016LayerNorm,He2016ResNet,Srivastava2014Dropout,PeeblesXie2023,Loshchilov2019AdamW,Loshchilov2017SGDR,vandenOord2016WaveNet,SchusterPaliwal1997,Cho2014GRU,Bouchacourt2016DISCO,Zerveas2021,Baum1970,GordonSalmondSmith1993,KongLiuWong1994,SchulzBarkaiMetzler2014,Feller1949,AsmussenGlynn2007,RizzoSzekely2010,Chernoff1952,Fisher1925,LanoiseleeData,Bellman1952,KatrukhaData,SittewelleRoyleData} therein) gives the training details with a comparison of alternative architectures, the mechanism-based reference distributions, and the assignment of single trajectories to processes. It also contains the derivations of Sec.~\ref{sec:origin}, the measurement under localization error, and the details of the measurement on experimental trajectories and of the segmentation.

\section{Local conditional information and its measurement}\label{sec:framework}

\begin{figure*}
\includegraphics[width=0.9\textwidth]{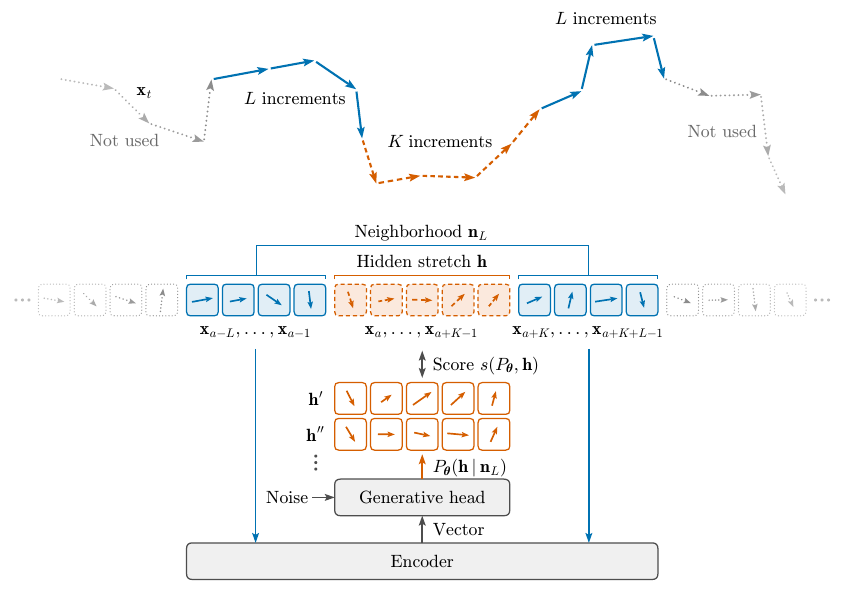}
\caption{Schematic diagram of a hidden stretch, its neighborhood, and masked learning. Top: a trajectory, drawn as one arrow per increment $\mathbf{x}_t$; the $K$ increments of the hidden stretch (dashed) are removed and predicted from the $L$ increments on either side (solid), while the rest of the trajectory (dotted) is not used. Middle: the same increments in time order, with the hidden stretch $\mathbf{h}$ and the two halves of the neighborhood $\mathbf{n}_L$ marked. Bottom: masked learning on these increments. The encoder reads the two halves of the neighborhood and compresses them into a vector. The generative head, conditioned on this vector, transforms random noise into samples $\mathbf{h}',\mathbf{h}'',\dots$ of the hidden stretch; the distribution of these samples is the prediction $P_{\boldsymbol{\theta}}(\mathbf{h}\mid\mathbf{n}_L)$. The recorded stretch $\mathbf{h}$ is the training target; after training, the prediction is scored against the recorded stretch by $s(P_{\boldsymbol{\theta}},\mathbf{h})$.}
\label{fig:intro}
\end{figure*}

This section presents the quantity that we measure and the way we measure it. Sec.~\ref{sec:information} defines the local conditional information and its second-order part. Sec.~\ref{sec:measurement} describes how masked learning estimates this information from trajectories alone and shows that the estimate is a lower bound, and Sec.~\ref{sec:settings} gives the model and the settings used in the main text.

\subsection{Local conditional information and its second-order part}\label{sec:information}

Here we define the local conditional information, which Sec.~\ref{introduction} introduced as a general description of the relation between increments, and then separate its second-order part. A trajectory is recorded as a sequence of {\it increments} $\mathbf{x}_1,\mathbf{x}_2,\dots,\mathbf{x}_T$, where $\mathbf{x}_t\in\mathbb{R}^d$ is the displacement during the $t$th sampling interval. As shown in Fig.~\ref{fig:intro}, a hidden stretch $\mathbf{h}$ is masked from the trajectory and predicted from its neighborhood $\mathbf{n}_L$. The stretch consists of $K$ consecutive increments starting at position $a$, $\mathbf{h}=(\mathbf{x}_a,\dots,\mathbf{x}_{a+K-1})\in\mathbb{R}^{dK}$. The neighborhood has two segments of equal length. One is the $L$ increments immediately before the stretch, $\mathbf{x}_{a-L},\dots,\mathbf{x}_{a-1}$; the other is the $L$ increments immediately after it, $\mathbf{x}_{a+K},\dots,\mathbf{x}_{a+K+L-1}$. Together they form $\mathbf{n}_L\in\mathbb{R}^{2dL}$. Here, $L = 0$ denotes an empty neighborhood.

A prediction of the hidden stretch is a conditional probability distribution $P(\mathbf{h}\mid\mathbf{n}_L)$: the joint distribution assigned to the $K$ hidden increments once the position $a$ and the neighborhood $\mathbf{n}_L$ are known. It need not have a probability density, and it may be available only through samples drawn from it. Here, masking supplies its own ground truth: the hidden stretch $\mathbf{h}$ was actually recorded, so every prediction can be checked against the real increments. We score a prediction by a scoring rule $s(P,\mathbf{h})$, a number that is low when the true stretch was predicted to be likely and high when it was predicted to be unlikely. Let $\langle\cdot\rangle$ denote the average over trajectories of the process and over the positions $a$ of the hidden stretch. The scoring rule is required to be strictly proper~\cite{GneitingRaftery2007}: the average score is lowest when the prediction equals the true distribution, and only then. Its lowest value over all predictions is \begin{equation} S(L)=\min_{P}\left\langle s(P,\mathbf{h})\right\rangle, \label{eq:S} \end{equation} where the minimum is taken over all predictions $P$ that use a neighborhood of length $L$. $S(L)$ does not depend on any particular prediction and measures how uncertain the hidden stretch remains once such a neighborhood is known. $S(L)$ also depends on the stretch length $K$, which is not written here.

The local conditional information is the amount by which a neighborhood of length $L$ lowers this uncertainty, \begin{equation} I(L)=S(0)-S(L), \label{eq:I} \end{equation} where $S(0)$ is the uncertainty of the hidden stretch when nothing around it is known; we call it the {\it prior score}. In terms of the score, $I(L)$ is the average gain of the best prediction that uses the neighborhood over the best prediction that does not. Since a longer neighborhood contains the shorter one, $S(L)$ cannot increase with $L$; $I(L)$ therefore starts from $I(0)=0$ and cannot decrease. It stays at zero only if the hidden stretch is statistically independent of its neighborhood, as in Brownian motion. Any dependence between the two, linear or nonlinear, makes the information positive. Nothing in this definition refers to how the trajectories were generated: it applies to any process, Gaussian or not, stationary or not. The curve $I(L)$ records how the information about the hidden stretch is distributed over its neighbors, and it is this curve that we measure for each process.

If the hidden stretch has a probability density, the logarithmic score~\cite{Good1952} can be used, $s(P,\mathbf{h})=-\ln p(\mathbf{h}\mid\mathbf{n}_L)$, the negative logarithm of the probability density $p$ that the prediction assigned to the ground truth. $S(L)$ is then the conditional entropy of the hidden stretch, and $I(L)$ is the mutual information between the hidden stretch and its neighborhood~\cite{Shannon1948}, in nats. For some of the processes studied here the hidden stretch has no probability density: during a waiting period all $K$ hidden increments are exactly zero, and within a single flight they repeat the increments next to them. The average logarithmic score then has no finite minimum, so it cannot measure the information, and the energy score~\cite{GneitingRaftery2007} takes its place, \begin{equation} s(P,\mathbf{h})=\overline{\|\mathbf{h}'-\mathbf{h}\|}-\frac{1}{2}\,\overline{\|\mathbf{h}'-\mathbf{h}''\|}, \label{eq:ES} \end{equation} where $\mathbf{h}'$ and $\mathbf{h}''$ are independent samples of the prediction $P$, the overline denotes the average over them, and $\|\cdot\|$ is the Euclidean norm in $\mathbb{R}^{dK}$. The energy score requires only samples of the prediction and remains strictly proper when no density exists. With it, the prior score $S(0)$ is half the mean distance between the hidden stretches of two independent trajectories at the same position. We use the energy score for all processes, so that their information is measured in the same way. Because this score carries the dimension of length, we express $I(L)$ and all quantities derived from it in units of the prior score $S(0)$ of each process; $I(L)=5\%$, for example, means that the neighborhood lowers the uncertainty of the hidden stretch by 5\%.

Part of the local conditional information is already described by second-order statistics, that is, by the mean and covariance of the increments, which determine the MSD and the VACF. To separate this part, we consider the Gaussian process whose increments have the same mean and covariance as those of the actual process. Its best prediction is kriging, which uses the covariance alone. We define $I_2(L)$ as the gain of kriging on the trajectories of the actual process, that is, the amount by which the average score of kriging with a neighborhood of length $L$ falls below that of kriging without a neighborhood. For the logarithmic score it is \begin{equation} I_2(L)=\frac{1}{2}\ln\frac{\det\mathbf{C}_{\mathbf{h}}\,\det\mathbf{C}_{\mathbf{n}}}{\det\mathbf{C}}, \label{eq:I2} \end{equation} where $\mathbf{C}_{\mathbf{h}}$, $\mathbf{C}_{\mathbf{n}}$, and $\mathbf{C}$ are the covariance matrices of the hidden stretch, of the neighborhood, and of the two taken together; for the energy score it is computed from the trajectories. We call $I_2(L)$ the {\it second-order part} of the local conditional information and write the remainder as \begin{equation} \Delta I(L)=I(L)-I_2(L). \label{eq:dI} \end{equation} For a Gaussian process such as FBM, $\Delta I(L)=0$ for every $L$, and the curve $I(L)$ is fixed by the correlation function. A positive $\Delta I(L)$ demonstrates a dependence between the hidden stretch and its neighborhood that the covariance does not describe.

\subsection{Measuring the local conditional information by masked learning}\label{sec:measurement}

To measure $I(L)$ one needs the prediction that attains the minimum in Eq.~(\ref{eq:S}), which is the true distribution of the hidden stretch conditioned on its neighborhood. For a Gaussian process, kriging gives this distribution in closed form. For a non-Gaussian process it does not, and if no generating mechanism is assumed, the distribution has to be learned from the trajectories themselves. Masked learning does this by making the prediction task of Sec.~\ref{sec:information} the training task. For each process, we train a conditional generative model, that is, a neural network with parameters $\boldsymbol{\theta}$ that receives a neighborhood $\mathbf{n}_L$ and generates samples of the hidden stretch (Fig.~\ref{fig:intro}, bottom). In each training example, a stretch of $K$ increments is masked from a trajectory of the process and the neighborhood length $L$ is drawn uniformly between 1 and 128. The sole training objective is to reconstruct this hidden stretch from the $L$ increments on either side, with the recorded increments as the target, so neither the generating mechanism nor any label enters the training. After training, the distribution of the samples that the model generates, $P_{\boldsymbol{\theta}}(\mathbf{h}\mid\mathbf{n}_L)$, is a prediction in the sense of Sec.~\ref{sec:information}, and one model provides it for every $L$ in this range. The average score of this prediction on test trajectories, $\hat{S}(L)$, takes the place of $S(L)$ in Eq.~(\ref{eq:I}) and gives our estimate of the local conditional information, \begin{equation} \hat{I}(L)=S(0)-\hat{S}(L). \label{eq:Ihat} \end{equation}

The trained model is one of the predictions over which the minimum in Eq.~(\ref{eq:S}) is taken, and the test trajectories take no part in its training. Its average score therefore cannot lie below the minimum, $\hat{S}(L)\ge S(L)$ up to the sampling error of the test set, with equality only if the model reproduces the true distribution. Hence $\hat{I}(L)\le I(L)$: the estimate is a lower bound on the true local conditional information, and every imperfection of the model, whether of its architecture or of its training, lowers it. Since $I_2(L)$ does not involve the model, the excess of the estimate over the second-order part is a lower bound on the part beyond second order, \begin{equation} \Delta\hat{I}(L)=\hat{I}(L)-I_2(L)\le\Delta I(L). \label{eq:bound} \end{equation} A positive $\Delta\hat{I}(L)$ therefore establishes that the neighborhood provides information beyond second order. This conclusion uses no property of the process and none of the network, so it holds for any generating mechanism and any architecture. A negative $\Delta\hat{I}(L)$ allows no conclusion: it means only that the shortfall of the estimate, $I(L)-\hat{I}(L)$, exceeds $\Delta I(L)$, as it does for any imperfect model of a Gaussian process.

\subsection{Model and settings}\label{sec:settings}

The masked generative model consists of an encoder and a generative head (Fig.~\ref{fig:intro}, bottom). The encoder reads the increments of the neighborhood, which it compresses into a vector. Conditioned on this vector, the generative head transforms random noise into a sample of the hidden stretch. Throughout the main text, the encoder is built from Transformer layers~\cite{Vaswani2017} and the generative head is a conditional flow trained by flow matching~\cite{Lipman2023FlowMatching,Liu2023RectifiedFlow}. The encoder has four layers of width 128 with four attention heads and rotary position embeddings~\cite{Su2024RoFormer}, and the network has about $1.5\times10^{6}$ parameters for $K=5$. A sample is drawn by integrating the flow in 50 Euler steps. The architecture and the loss are the same for every process in the main text. Only the scale of the noise fed to the flow is set separately for each process. Each model is trained on $10^{5}$ trajectories, and the weights saved at the end of training are used. Sec.~S1 of the SM gives the training details and compares alternative encoders and generative heads.

Five of the processes are the canonical mechanisms of the AnDi challenge~\cite{AnDi2021}: FBM, SBM, the continuous-time random walk (CTRW)~\cite{MontrollWeiss1965,ScherMontroll1975}, the annealed transient time motion (ATTM)~\cite{Massignan2014}, and the L\'evy walk (LW)~\cite{ShlesingerWestKlafter1987,KlafterBlumenShlesinger1987}. They are grouped by MSD exponent: $\alpha=0.4$ (SBM, FBM, ATTM \footnote{For the ATTM, the MSD exponent alone does not uniquely characterize its dynamics, which are governed by two parameters, $\sigma$ and $\gamma$ (SM, Sec.~S2~B). Here $\alpha$ specifies the parameterization $\sigma=2\alpha$ and $\gamma=2$, yielding asymptotic MSD exponents of $0.7$ for $\alpha=0.7$ and $0.5$ for $\alpha=0.4$~\cite{Massignan2014}.}, and CTRW), $\alpha=0.7$ (FBM, ATTM, and CTRW), $\alpha=1.3$ (FBM and LW), and $\alpha=1.6$ (SBM, FBM, and LW), with Hurst exponent $H=\alpha/2$ for FBM. Brownian motion is included as a reference. The two additional processes are a two-state process and a diffusing-diffusivity (DD) process, both Brownian yet non-Gaussian~\cite{Wang2009,Chechkin2017}. In both, each increment is Gaussian with zero mean and variance $2D_t$ per coordinate, where the diffusivity $D_t$ fluctuates in time, so that the MSD is linear and the increments are uncorrelated. They differ in how $D_t$ fluctuates. In the two-state process~\cite{Karger1969}, $D_t$ switches between 0.1 and 1 with probability 0.01 per step. In the DD process~\cite{ChubynskySlater2014,Chechkin2017}, $D_t$ is the square of an Ornstein--Uhlenbeck process~\cite{UhlenbeckOrnstein1930} with a correlation time of 100 steps, discretized into 48 states and normalized to unit mean. Both processes start in the stationary state of $D_t$. Taken together, the processes cover Gaussian and non-Gaussian statistics, correlated and uncorrelated increments, and stationary and nonstationary dynamics.

The trajectories of all processes are two-dimensional, $d=2$, and consist of $T=499$ increments. The energy score of the model is evaluated from 512 samples for each neighborhood. Unless stated otherwise, $K=5$, and the averages are taken over 2000 test trajectories of each process and over 12 positions of the hidden stretch. The positions span the range in which 128 increments are available on either side of the stretch, from $a=129$ to $a=367$ for $K=5$, and $\hat{I}(L)$ is evaluated at $L=1,2,4,\dots,128$. Kriging uses the exact covariance of each process. Quoted intervals are 95\% bootstrap intervals~\cite{Efron1979} obtained by resampling whole trajectories. Any departure from the settings of this subsection is stated where it occurs.

\section{Local conditional information of diffusion processes}\label{sec:results}

\begin{figure*}[t]
\includegraphics[width=0.95\textwidth]{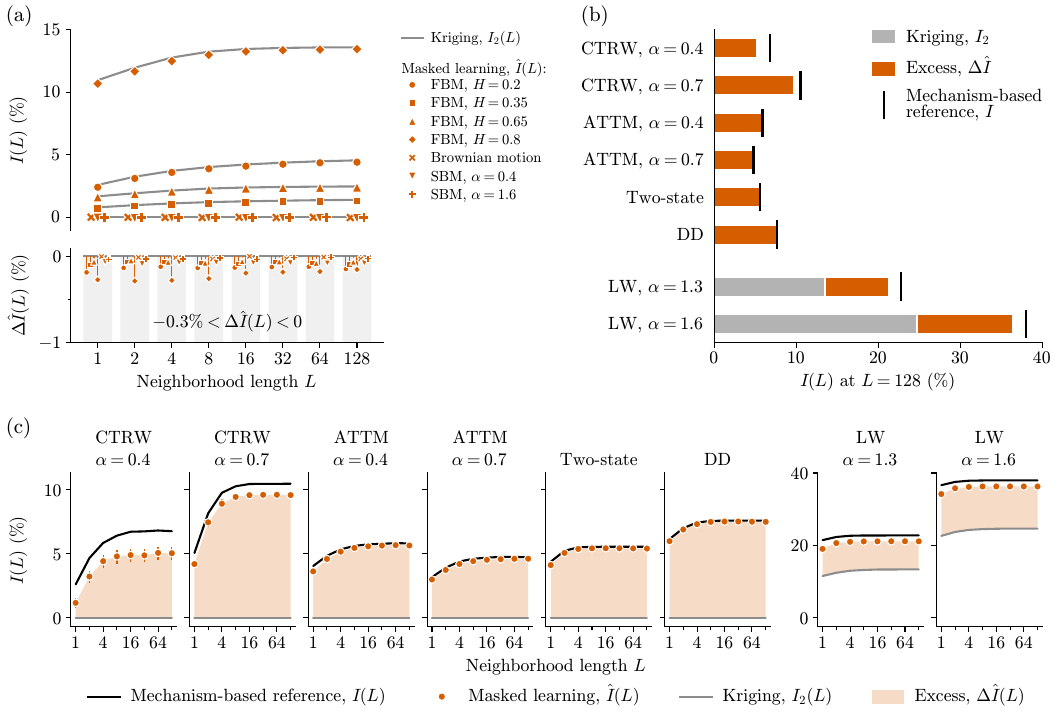}
\caption{Local conditional information estimated by masked learning, compared with kriging and with the mechanism-based reference. All quantities are expressed in percent of the prior score $S(0)$, and the neighborhood length $L$ is shown on a logarithmic scale. (a) Gaussian processes. Top: the gain of kriging, $I_2(L)$ (gray lines), and the gain of masked learning, $\hat{I}(L)$ (symbols). Bottom: their difference $\Delta\hat{I}(L)$ [Eq.~(\ref{eq:bound})]; the gray line marks zero, and a thin line connects each symbol to it. Symbols that would overlap at the same $L$ are displaced horizontally. In the bottom panel they follow the order of the legend, and each light-gray band contains the symbols of one $L$. (b) Non-Gaussian processes at $L=128$. The full length of each bar is $\hat{I}$: the gray part is the gain of kriging, $I_2$, and the vermilion part is the excess $\Delta\hat{I}$. The black vertical line marks the gain of the mechanism-based reference, $I$. The upper six processes have uncorrelated increments, so that $I_2=0$. (c) The same quantities as functions of $L$ for the processes in (b): the reference $I(L)$ (black lines), masked learning $\hat{I}(L)$ (circles), kriging $I_2(L)$ (gray lines), and the excess $\Delta\hat{I}(L)$ (shaded area). Error bars are 95\% bootstrap intervals and are mostly smaller than the symbols.}
\label{fig:gain}
\end{figure*}

This section reports the measurement of the local conditional information by masked learning, first for the processes of Sec.~\ref{sec:settings} and then for experimental trajectories. Sec.~\ref{sec:gaussian} begins with the Gaussian processes, for which kriging gives the information exactly, and shows that the estimate reproduces it. Sec.~\ref{sec:nongaussian} turns to the non-Gaussian processes, where the estimate exceeds the gain of kriging, and compares the excess with the information beyond second order that follows from their generating mechanisms. Sec.~\ref{sec:identical} considers processes with identical second-order statistics and shows that the local conditional information allows single trajectories to be assigned to the process that generated them. Sec.~\ref{sec:origin} presents a general theory that reveals the origin of the information beyond second order. Finally, Sec.~\ref{sec:experiment} applies the measurement directly to experimental trajectories, with no mechanism assumed.

\subsection{Gaussian processes}\label{sec:gaussian}

For a Gaussian process, the best prediction of a hidden stretch from its neighborhood is known exactly, and with it the local conditional information. The distribution of the hidden stretch conditioned on its neighborhood is again Gaussian. For increments of zero mean, as in all processes considered here, the mean and covariance of this distribution are \begin{equation} \begin{aligned} \boldsymbol{\mu}&=\mathbf{C}_{\mathbf{h}\mathbf{n}}\mathbf{C}_{\mathbf{n}}^{-1}\mathbf{n}_L,\\ \boldsymbol{\Sigma}&=\mathbf{C}_{\mathbf{h}}-\mathbf{C}_{\mathbf{h}\mathbf{n}}\mathbf{C}_{\mathbf{n}}^{-1}\mathbf{C}_{\mathbf{h}\mathbf{n}}^{\top}. \end{aligned} \label{eq:kriging} \end{equation} Here $\mathbf{C}_{\mathbf{h}}$ and $\mathbf{C}_{\mathbf{n}}$ are the covariance matrices in Eq.~(\ref{eq:I2}), and $\mathbf{C}_{\mathbf{h}\mathbf{n}}$ is the covariance between the hidden stretch and the neighborhood. This distribution is the kriging prediction used in Sec.~\ref{sec:information} to define $I_2(L)$. For a Gaussian process kriging is optimal: it is the true conditional distribution and attains the minimum in Eq.~(\ref{eq:S}), so that no prediction based on the same neighborhood can score better and $I(L)=I_2(L)$. Among the processes in Sec.~\ref{sec:settings}, Brownian motion, SBM, and FBM are Gaussian. The increments of Brownian motion and of SBM are independent, so that $\mathbf{C}_{\mathbf{h}\mathbf{n}}=\mathbf{0}$: the prediction does not depend on the neighborhood, and $I(L)=0$. The increments of FBM are correlated, and kriging profits from the neighborhood: at $L=128$ its gain lies between 1.4\% ($H=0.35$) and 13.6\% ($H=0.8$) for the four Hurst exponents in Sec.~\ref{sec:settings}. For each of them the gain rises steeply at small $L$ and grows only slowly thereafter [gray lines in Fig.~\ref{fig:gain}(a)].

Masked learning predicts the hidden stretch as accurately as kriging [Fig.~\ref{fig:gain}(a)]. In the top panel, the symbols of masked learning follow the kriging curve of each process over the whole range of $L$. At every neighborhood length, its gain $\hat{I}(L)$ stays below that of kriging, which is optimal, and the difference is less than 0.3\% of the prior score. The bottom panel shows this difference as $\Delta\hat{I}(L)$ of Eq.~(\ref{eq:bound}); its magnitude is largest for FBM with $H=0.8$, the process with the largest gain, and stays below 0.2\% for all other processes. For Brownian motion and SBM, where the neighborhood cannot help, $\hat{I}(L)$ lies between $-0.09\%$ and $0.00\%$ at every $L$. Masked learning thus attains, from the trajectories alone, the accuracy that kriging reaches with the exact covariance. Since it nowhere surpasses kriging, the measurement shows no information beyond second order in the Gaussian processes.

\subsection{Non-Gaussian processes}\label{sec:nongaussian}

For the non-Gaussian processes, masked learning predicts the hidden stretch more accurately than kriging. Kriging is no longer the true conditional distribution of such a process, and its gain can be surpassed. The two gains at $L=128$ are compared in Fig.~\ref{fig:gain}(b). Each process has one bar, whose full length is the gain of masked learning and whose gray part is the gain of kriging. The remaining part is the excess $\Delta\hat{I}(L)$ of the gain of masked learning over that of kriging. It is positive for every process and lies between 4.6\% and 11.6\% of the prior score. The same gains are shown as functions of $L$ in Fig.~\ref{fig:gain}(c). The excess is also positive at every shorter neighborhood, down to a single increment on either side. It is present whether or not kriging itself profits from the neighborhood. In the LW, whose increments are correlated because a flight keeps its velocity over several steps, masked learning adds the excess to a gain that kriging already achieves, and about one third of the gain of masked learning is excess. In the CTRW, the ATTM, and the two-state and DD processes, whose increments are uncorrelated, kriging gains nothing, and the whole gain of masked learning is excess. Kriging uses everything that second-order statistics provide. The excess therefore shows that in these processes the neighbors constrain the hidden stretch through relations that the covariance does not describe. Since no prediction can outperform the true conditional distribution, the information beyond second order is at least as large as the excess.

How close this excess comes to the information beyond second order can be determined for these processes, because their generating mechanisms are known. From the mechanism we compute the true conditional distribution of the hidden stretch, which we call the {\it mechanism-based reference}. The computation uses renewal theory for the CTRW and the LW and inference of the hidden diffusivity for the other processes (SM, Sec.~S2). The gain of the reference is the local conditional information $I(L)$ itself [black lines in Figs.~\ref{fig:gain}(b) and \ref{fig:gain}(c)], and its excess over kriging is $\Delta I(L)$. At every neighborhood length the gain of masked learning stays below that of the reference [Fig.~\ref{fig:gain}(c)], and at $L=128$ the excess $\Delta\hat{I}(L)$ amounts to between 75\% and 99\% of $\Delta I(L)$. This fraction is highest, 97\% to 99\%, for the ATTM and for the two-state and DD processes, whose bars in Fig.~\ref{fig:gain}(b) end at the reference, and lowest for the CTRW with $\alpha=0.4$. The gain of masked learning also follows the reference in its dependence on $L$: in every process the two curves rise over the same neighborhood lengths and level off together. The reference requires the mechanism and its parameters, whereas masked learning uses the trajectories alone. The measured excess is thus a lower bound that captures most of the information beyond second order. Taken together with Sec.~\ref{sec:gaussian}, these results show how widely processes with the same MSD exponent differ in their local conditional information. At $\alpha=1.6$ and $L=128$, the estimate is zero for SBM, 13.4\% of the prior score for FBM, and 36.3\% for the LW [Figs.~\ref{fig:gain}(a) and \ref{fig:gain}(b)].

\subsection{Processes with identical second-order statistics}\label{sec:identical}

\begin{figure*}[t]
\includegraphics[width=0.95\textwidth]{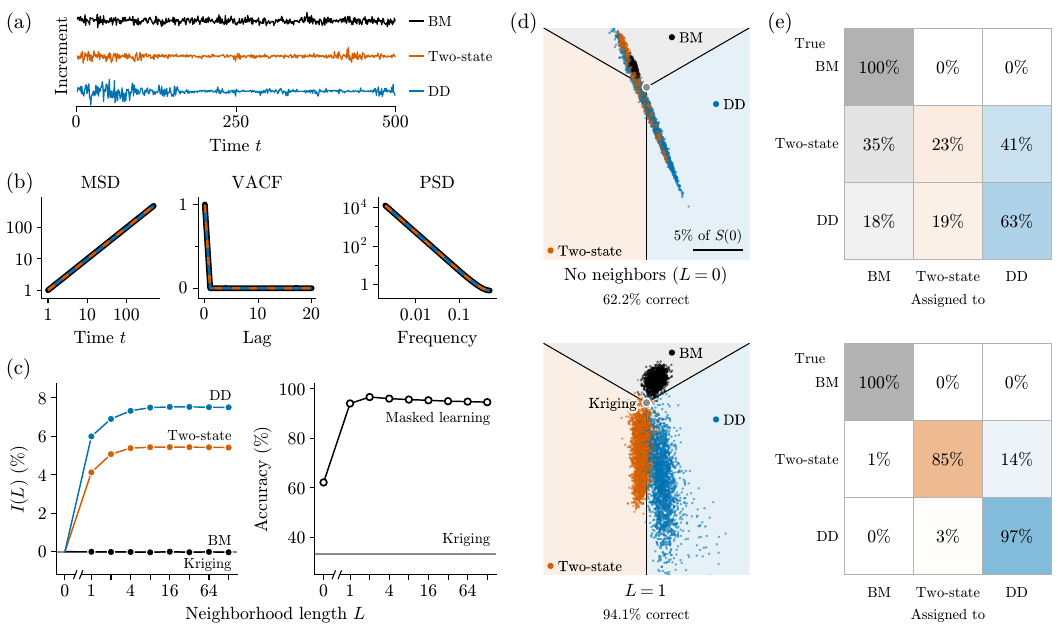}
\caption{Local conditional information of processes with identical second-order statistics: Brownian motion (BM) and the two-state and DD processes. (a) Increments (one coordinate) of one trajectory of each process, on the same scale. (b) MSD, VACF, and PSD of the three processes; the curve of BM is drawn wide and those of the other two processes narrow, inside it. (c) Left: gain of kriging, $I_2(L)$ (gray line), and gain of masked learning, $\hat{I}(L)$ (filled circles), from the data of Fig.~\ref{fig:gain}. Right: accuracy, the percentage of the test trajectories assigned to the process that generated them, with kriging (gray line) and with masked learning (open circles). At $L=0$ no neighbors are used. (d) Test trajectories scored under the predictions of the three processes, without neighbors (top) and with $L=1$ (bottom). Each point is one trajectory, colored by the process that generated it and placed according to the differences between its three scores. A score lower by 5\% of $S(0)$ under one prediction displaces the point by the length of the bar. The gray dot marks kriging. Points outside the frames are not shown. (e) Percentage of the trajectories of each process (rows) assigned to each process (columns), for the two cases of (d).}
\label{fig:identical}
\end{figure*}

Secs.~\ref{sec:gaussian} and \ref{sec:nongaussian} showed that the prediction learned by masked learning matches kriging in the Gaussian processes and surpasses it in the non-Gaussian ones, where it captures what second-order statistics do not describe. This information beyond second order matters most for processes whose second-order statistics are identical. Because their MSDs, VACFs, and PSDs coincide, second-order analysis cannot distinguish them, and only what lies beyond second order can. We call such processes degenerate at second order. Brownian motion and the two-state and DD processes form such a group: their increments are uncorrelated, with a variance that is constant along the trajectory. Their covariances thus differ only by the unit of length, which we set equal for the three processes (SM, Sec.~S3). The fluctuating diffusivity of the two-state and DD processes is visible in single trajectories [Fig.~\ref{fig:identical}(a)] but not in the MSD, the VACF, or the PSD [Fig.~\ref{fig:identical}(b)]. Kriging gives the same prediction for the three processes, and its gain vanishes, $I_2(L)=0$ [Fig.~\ref{fig:identical}(c), left]. The gain of masked learning, by contrast, vanishes for Brownian motion and reaches 5.4\% and 7.5\% of the prior score at $L=128$ for the two-state and DD processes. Here we test whether a single trajectory can be assigned to the process that generated it, and how much of this identification is due to the local conditional information.

For each process, masked learning has provided a prediction learned from the trajectories of that process alone. A test trajectory is scored under the predictions of the three processes and is assigned to the process whose prediction attains the lowest energy score, averaged over the hidden stretches of the trajectory (SM, Sec.~S3). We use 48 adjacent hidden stretches, $a=129,134,\dots,364$. The right part of Fig.~\ref{fig:identical}(c) shows the accuracy of this assignment against the neighborhood length. In Fig.~\ref{fig:identical}(d), each test trajectory is a point whose position is set by the differences between its three scores: it is displaced from the center toward the process whose prediction scores it best (SM, Sec.~S3). The percentage of the trajectories of each process that are assigned to each of the three is given in Fig.~\ref{fig:identical}(e). With kriging the three scores are equal: every trajectory lies at the center [gray dot in Fig.~\ref{fig:identical}(d)], and the assignment becomes a random choice, correct for one third of the trajectories [gray line in Fig.~\ref{fig:identical}(c), right].

With masked learning, the accuracy depends on the length of the neighborhood [Fig.~\ref{fig:identical}(c), right]. Without neighbors, each prediction is the distribution of the hidden stretch of the same process, the best prediction for an empty neighborhood. The assignment is then correct for 62.2\% of the trajectories [top panels of Figs.~\ref{fig:identical}(d) and \ref{fig:identical}(e)]. Every Brownian trajectory is recognized, but 35\% of the two-state and 18\% of the DD trajectories are taken for Brownian motion, and the two non-Gaussian processes are hardly separated from each other. An assignment between these two alone is correct for 61.0\% of their trajectories. A single increment on either side of each hidden stretch, $L=1$, raises the accuracy from 62.2\% to 94.1\% [bottom panels of Figs.~\ref{fig:identical}(d) and \ref{fig:identical}(e)]. Only 19 of the 4000 trajectories of the two non-Gaussian processes are still taken for Brownian motion. An assignment between these two processes alone is now correct for 91.5\% of their trajectories. Longer neighborhoods change the accuracy little: up to $L=128$ it remains between 94.1\% and 96.7\%. The local conditional information rises in the same way [Fig.~\ref{fig:identical}(c), left]: the nearest neighbors provide more than three quarters of its value at $L=128$. With $L=1$, a two-state trajectory is scored better by the prediction of its own process than by that of Brownian motion by 5.2\% of $S(0)$ on average, and a DD trajectory by 8.8\%. Of these differences, 4.1\% and 6.0\% are the local conditional information, which the prediction of Brownian motion does not use. The remainder, 1.1\% and 2.8\%, stems from the non-Gaussian distribution of the hidden stretch itself. A single stretch assigns only half of the trajectories correctly, and the accuracy rises as the scores of the 48 stretches are averaged. Most of the errors that remain are two-state trajectories assigned to the DD process, 14\% of them [Fig.~\ref{fig:identical}(e), bottom]. On two-state trajectories, the score of the DD prediction is higher than that of the two-state prediction by only 0.6\% of $S(0)$ on average.

These results show that identical second-order statistics do not make processes indistinguishable. Second-order statistics fix only the mean and covariance of the increments, whereas masked learning learns the full conditional distribution of a hidden stretch. Under a strictly proper score, the trajectories of a process are scored best on average by its own conditional distribution and strictly worse by that of any process for which this distribution differs. This holds whether or not the second-order statistics of the two processes coincide, and the difference accumulates over the hidden stretches of a trajectory. For the two-state and DD processes, most of this difference from Brownian motion is the local conditional information measured in Sec.~\ref{sec:nongaussian}. The nearest neighbors carry most of it. This information therefore removes the degeneracy that limits every second-order analysis: processes with identical second-order statistics can be identified from single trajectories, with no input other than trajectories of each process. What limits the identification is then not the coincidence of second-order statistics but only the similarity of the conditional distributions and the length of the trajectory.

\subsection{Theoretical origin of the local conditional information}\label{sec:origin}

\begin{figure*}[t]
\includegraphics[width=1.0\textwidth]{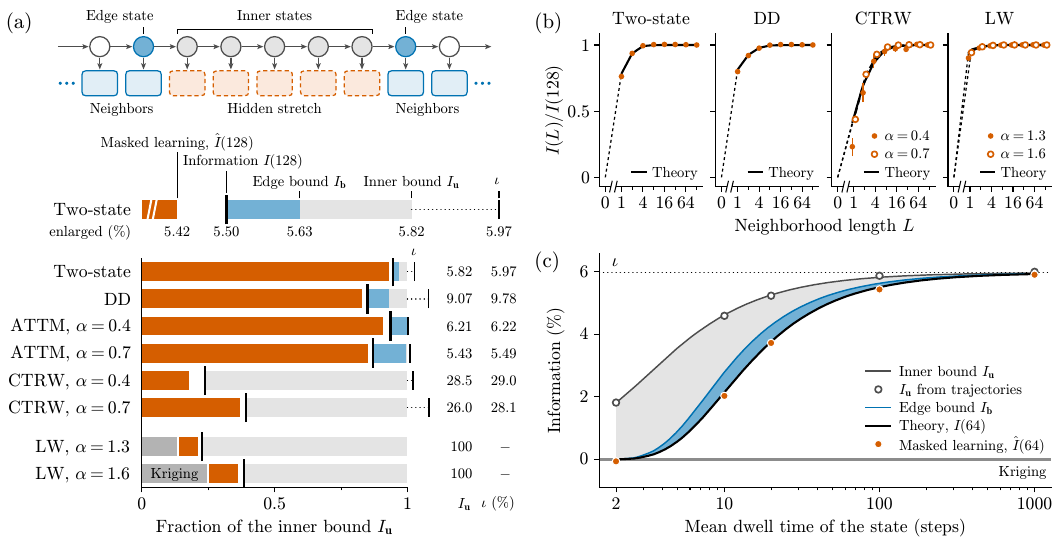}
\caption{Theoretical origin of the local conditional information. (a) Top: state model of Eq.~(\ref{eq:state}). Each state generates one increment. The edge states (blue) adjoin the hidden stretch, and the inner states (gray) belong to it. Middle: right end of the bar of the two-state process, enlarged, in percent of the prior score $S(0)$. Bottom: bounds and measured information at $L=128$ as fractions of the inner bound $I_{\mathbf{u}}$. Vermilion bars show masked learning, and their gray part is the kriging gain. Black lines mark the theoretical value of $I(128)$ (SM, Sec.~S4); for the ATTM they mark the mechanism-based reference on the test trajectories. The blue part ends at the edge bound $I_{\mathbf{b}}$, which coincides with $I(128)$ for the CTRW and the LW and lies within 0.5\% of $I_{\mathbf{u}}$ for the ATTM. The light gray part ends at $I_{\mathbf{u}}$. Thin ticks at or beyond the right end of the bars mark the theoretical value $\iota$ of Eq.~(\ref{eq:iota}) for the processes with a diffusivity state. The two columns on the right list $I_{\mathbf{u}}$ and $\iota$ in percent of $S(0)$. (b) Share $I(L)/I(128)$ of the information that a neighborhood of length $L$ provides. The black lines are theoretical (population averages; SM, Sec.~S4) and contain no adjustable parameters, and circles show masked learning. The theoretical values at $L=1,2,4,\dots,128$ are joined by straight lines, and the dotted segments between $L=0$ and $L=1$ start from $I(0)=0$. Dashed lines and open circles refer to the larger $\alpha$. (c) Two-state process with different mean dwell times of the state. The black line is $I(64)$, and the filled circles are the estimates of masked learning. The gray and blue lines are the inner and edge bounds. All lines are theoretical (SM, Sec.~S4). The open circles are the inner bound evaluated with Eq.~(S20) of the SM on test trajectories, and the dotted line is $\iota$ of Eq.~(\ref{eq:iota}). The gray horizontal line is the kriging gain, which is zero. Error bars are 95\% bootstrap intervals.}
\label{fig:theory}
\end{figure*}

Secs.~\ref{sec:gaussian}--\ref{sec:identical} measured the local conditional information process by process. We now turn to its theoretical origin, which can be derived without reference to a particular mechanism. For any stochastic process, the local conditional information equals one quarter of the mean energy distance~\cite{BaringhausFranz2004,SzekelyRizzo2013} between the conditional distributions of the hidden stretch that belong to two independent neighborhoods [SM, Eq.~(S16)]. It thus measures how much this conditional distribution varies from one neighborhood to another. In a Gaussian process only its mean varies, linearly with the neighboring increments [Eq.~(\ref{eq:kriging})], and $I(L)$ is the kriging gain. Information beyond second order can arise only when the conditional distribution differs from the kriging prediction. A state that persists in time has this effect.

We consider processes whose increments are governed by a state $z_t$ of the particle or of its surroundings, such as a diffusivity, a velocity, or a mode of motion, \begin{equation} \mathbf{x}_t=\mathbf{f}(z_t,\boldsymbol{\xi}_t). \label{eq:state} \end{equation} Here $z_t$ is the state during the $t$th sampling interval, the $\boldsymbol{\xi}_t$ are random variables that are independent of one another and of the states, and $\mathbf{f}$ is an arbitrary function. The states form a Markov chain and may be discrete or continuous. Processes of this form are known as hidden Markov models or state-space models~\cite{BaumPetrie1966,CappeMoulinesRyden2005}. They include diffusion with a switching or a fluctuating diffusivity, motion in alternating modes, and, with the elapsed time counted as part of the state, renewal processes such as the CTRW and the LW. Since the increments are independent once the states are known and the states are Markovian, the hidden stretch depends on its neighborhood only through the two states adjacent to it, $\mathbf{b}=(z_{a-1},z_{a+K})$, which we call the edge states [Fig.~\ref{fig:theory}(a), top]. Its conditional distribution is a mixture of the distributions $P_{\mathbf{b}}$ of the hidden stretch for known edge states, weighted by the probability $p(\mathbf{b}\mid\mathbf{n}_L)$ that the neighborhood assigns to them. All that the neighbors can do is to infer the state at the two edges of the stretch. The information that this inference provides is (SM, Sec.~S4) \begin{equation} I(L)=\frac{1}{4}\Big[\big\langle\mathcal{E}(P_{\mathbf{b}},P_{\mathbf{b}'})\big\rangle_{0}-\big\langle\mathcal{E}(P_{\mathbf{b}},P_{\mathbf{b}'})\big\rangle_{L}\Big], \label{eq:origin} \end{equation} where $\mathcal{E}$ denotes the energy distance. The first average runs over the edge states $\mathbf{b}$ and $\mathbf{b}'$ of two independent trajectories, and the second over two independent draws from the posterior $p(\mathbf{b}\mid\mathbf{n}_L)$ of the same neighborhood. The first term is the information $I_{\mathbf{b}}$ that exactly known edge states would provide, and no observation outside the stretch can provide more, $\hat{I}(L)\le I(L)\le I_{\mathbf{b}}$. The second term is the part that $L$ neighbors leave unresolved.

Eq.~(\ref{eq:state}) also determines what second-order statistics register of the state. Because the hidden stretch and the neighborhood are independent once the states are known, their covariance equals the covariance of their conditional means [SM, Eq.~(S19)]. If the state leaves the mean of the increments at zero, as a diffusivity does, this covariance vanishes however long the state persists. Kriging then gains nothing, and all of $I(L)$ lies beyond second order: the neighbors change the width of the conditional distribution, not its mean. If the state sets the mean of the increments, as a velocity does, kriging has a gain. An excess can remain, because the uncertainty of the stretch depends on its neighborhood, whereas the covariance of the kriging prediction does not [Eq.~(\ref{eq:kriging})]. This is the pattern of Fig.~\ref{fig:gain}(b): the state is a diffusivity in the two-state, DD, ATTM, and CTRW processes and a velocity in the LW.

The distribution of the state sets the size of the information. For Gaussian increments of zero mean and variance $2D(z_t)$ per coordinate, and a state that does not change over the stretch and its neighborhood, the edge bound is \begin{equation} \frac{I_{\mathbf{b}}}{S(0)}=\iota\equiv1-\frac{\langle\sqrt{2D}\rangle}{\langle\sqrt{D+D'}\rangle}, \label{eq:iota} \end{equation} where $D$ and $D'$ are two independent values of the diffusivity. For weak heterogeneity $\iota$ tends to $A/16$, where $A=\langle D^2\rangle/\langle D\rangle^2-1$ is the non-Gaussian parameter of a single increment~\cite{Rahman1964}. For the non-Gaussian processes of Fig.~\ref{fig:gain}(b), Fig.~\ref{fig:theory}(a) compares the bounds with the measured information. Each bar shows the estimate of masked learning, the information $I(128)$, and the edge bound as fractions of the inner bound $I_{\mathbf{u}}$. The inner bound is the information that exactly known states inside the stretch would provide (SM, Sec.~S4). For the two-state process, $\iota=5.97\%$. The switching of the state lowers the edge bound to 5.63\%, the information at $L=128$ is 5.50\%, and masked learning recovers 5.42\% [Fig.~\ref{fig:theory}(a), middle]. For the DD process, whose diffusivity is the square of a Gaussian variable, $\iota=9.78\%$ and the information is 7.69\%: the diffusivity changes within the stretch, and a continuous state is never determined exactly. The share $I(L)/I(128)$ of the information that a neighborhood of length $L$ provides is shown in Fig.~\ref{fig:theory}(b), from the theory (lines) and from masked learning (circles). A state with two values is resolved by few neighbors: in the two-state process, a single neighbor on either side provides 78\% of $I(128)$, and the measured share is 76\%. This is the origin of the step between $L=0$ and $L=1$ in Fig.~\ref{fig:identical}(c). The other processes are treated in the SM (Sec.~S4).

How much of this information the neighbors provide depends on the persistence of the state, which the distribution of a single increment does not show. If the state is drawn independently at every step, Eq.~(\ref{eq:state}) makes the hidden stretch independent of its neighborhood, and $I(L)=0$ for any distribution of the increments. In the two-state process, such a redrawing leaves the distribution of each increment unchanged, and with it $\iota$, $A$, and all second-order statistics. As the mean dwell time of the state increases from 2 to 1000 steps, the information at $L=64$ rises from zero to 5.92\% and approaches $\iota$ [Fig.~\ref{fig:theory}(c)]. Masked learning, applied to trajectories with mean dwell times of 2, 10, 20, and 1000 steps, reproduces this rise to within 0.12 percentage points. For a state that acts on the width of the increments, the local conditional information is thus the part of the non-Gaussianity that persists in time and can be inferred from the neighbors.

\subsection{Experimental trajectories}\label{sec:experiment}

\begin{figure*}[t]
\includegraphics[width=1.0\textwidth]{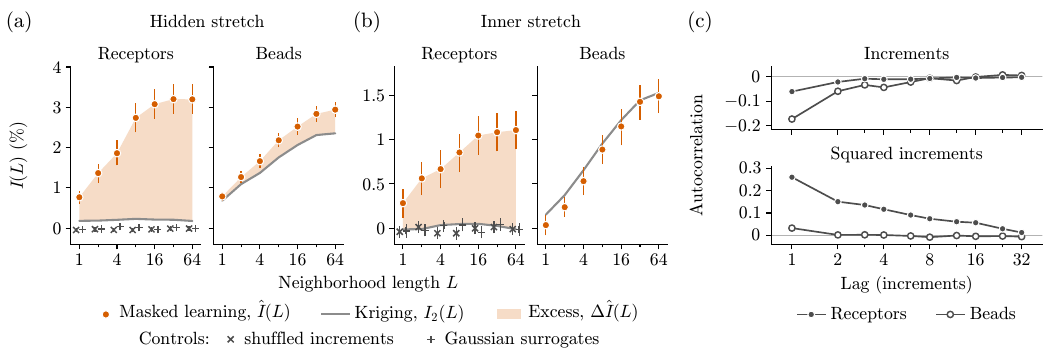}
\caption{Local conditional information of experimental trajectories: $\beta_2$-adrenergic receptors in the membrane of living cells~\cite{Lanoiselee2021} and tracer beads in an F-actin network~\cite{Granik2019}. For each system, the model is trained on the experimental trajectories themselves, and kriging uses their estimated covariance; $K=5$. All gains are in percent of the prior score of the stretch that is scored. (a) Hidden stretch: gain of masked learning, $\hat{I}(L)$ (circles), gain of kriging, $I_2(L)$ (gray lines), and the excess $\Delta\hat{I}(L)$ (shaded area). Crosses and plus signs show $\hat{I}(L)$ for two controls of the receptors, displaced horizontally: windows with shuffled increments, and Gaussian surrogates with the power spectrum of the increments of each window. (b) The same quantities for the inner stretch, the three inner increments of the hidden stretch. (c) Autocorrelation of the increments (top) and of the squared increments $\|\mathbf{x}_t\|^{2}$ (bottom) on the test windows, for the receptors (filled circles) and the beads (open circles). Error bars are 95\% bootstrap intervals obtained by resampling tracks; in (c) they are mostly smaller than the symbols.}
\label{fig:expinfo}
\end{figure*}

Secs.~\ref{sec:gaussian}--\ref{sec:origin} measured the local conditional information on trajectories simulated from known mechanisms. As a general measure of the relation between increments, it is defined for any process and should therefore be measurable on experimental trajectories as well. We now measure it for two systems: single $\beta_2$-adrenergic receptors in the plasma membrane of living cells, recorded at 30~ms per frame~\cite{Lanoiselee2021}, and tracer beads in an F-actin network~\cite{Granik2019}. From tracks of at least 200 positions we take windows of 200 positions and divide each window by its root-mean-square increment, so that differences in mobility between tracks do not enter the measurement. The tracks are split into a training, a validation, and a test set. For each system, the network of Sec.~\ref{sec:settings} is trained by masked learning on the windows of the training tracks alone, 3270 tracks of the receptors and 800 of the beads, with $L$ up to 64. These windows contain $8.4\times10^{5}$ and $1.4\times10^{6}$ increments, against $5\times10^{7}$ in each training set of Sec.~\ref{sec:settings}. The gains are evaluated on 756 windows from the 727 test tracks of the receptors and on 1000 windows from the 154 test tracks of the beads. Kriging uses the stationary covariance estimated from the training windows (SM, Sec.~S6).

Fig.~\ref{fig:expinfo}(a) shows the gain of masked learning, $\hat{I}(L)$, and the gain of kriging, $I_2(L)$, on the hidden stretch of each system. For the receptors, the gain of masked learning rises with the neighborhood length and levels off near $L=16$, about half a second on either side of the stretch. The gain of kriging stays close to zero. At $L=64$, $\hat{I}=3.2\%$ of the prior score and $I_2=0.2\%$, which leaves an excess of $\Delta\hat{I}=3.0\%$, with a 95\% interval from 2.7\% to 3.4\%. Nearly all of this information lies beyond second order. For the beads, the two gains rise together: at $L=64$, masked learning gains 2.9\% and kriging 2.4\%. The crosses and plus signs in Fig.~\ref{fig:expinfo}(a) are two controls for the receptors, each with a model of its own. One consists of windows whose increments are shuffled, the other of Gaussian surrogates that keep the power spectrum of the increments of each window~\cite{Theiler1992}. In both, the gain of masked learning vanishes. The gain of the receptors therefore stems from the order of their increments and not from their power spectrum.

Experimental positions carry a localization error~\cite{SavinDoyle2005,Berglund2010,Michalet2010}, and successive increments share the error of the position between them. The first and the last position of the hidden stretch also enter its neighbors, so that its two outer increments are partly predictable whatever the motion of the particle (SM, Sec.~S5). Fig.~\ref{fig:expinfo}(b) therefore shows the same gains for the three inner increments alone, the {\it inner stretch}. None of its four positions enters an increment of the neighborhood, and the shared errors do not contribute to its gain. For the receptors the excess remains: at $L=64$, $\Delta\hat{I}=1.1\%$ of the prior score of the inner stretch (0.9\% to 1.4\%), while the gain of kriging is zero. Recordings with the agonist isoproterenol and with the antagonist propranolol give the same excess on the inner stretch within the intervals (SM, Table~S5). The two controls show no gain on either stretch, $|\hat{I}(L)|\le0.06\%$ at every $L$. For the beads, masked learning and kriging coincide on the inner stretch, so that the small excess of the beads on the hidden stretch belongs to the two outer increments. The excess is a lower bound on the information beyond second order [Eq.~(\ref{eq:bound})]. With no mechanism assumed, the measurement thus establishes such information in the trajectories of the receptors, whereas on the inner stretch of the beads it shows none.

In the theory of Sec.~\ref{sec:origin}, information without a gain of kriging is the signature of a state that acts on the width of the increments and persists in time. Fig.~\ref{fig:expinfo}(c) examines both properties with two statistics that involve no model: the autocorrelation of the increments and that of the squared increments $\|\mathbf{x}_t\|^{2}$. The increments of the receptors are nearly uncorrelated, with a correlation of $-0.06$ at a lag of one increment. Their squared increments, by contrast, stay correlated over tens of increments: the correlation is 0.15 at a lag of two increments, 0.06 at 16, and 0.01 at 32. This is the range of neighborhood lengths over which $\hat{I}(L)$ grows. The beads show the opposite: their increments are anticorrelated, with $-0.17$ at a lag of one increment, which is what kriging uses, and the correlation of their squared increments does not exceed 0.03. The receptors behave as expected for a diffusivity state. For such a state, the information cannot exceed the inner bound of Sec.~\ref{sec:origin}, which measures the non-Gaussianity of the stretch and follows from the trajectories alone [SM, Eq.~(S20)]. On the inner stretch of the receptors it amounts to 3.6\%, so that the neighbors account for about 30\% of this non-Gaussianity. Shuffling leaves the distribution of the increments unchanged and removes the information, as the redrawing of the state does in Sec.~\ref{sec:origin}. The information of the receptors is thus the part of their non-Gaussianity that persists in time, consistent with the transient trapping detected in the same trajectories~\cite{Lanoiselee2021}.

\section{Unsupervised trajectory segmentation based on local conditional information}\label{sec:segmentation}

\begin{figure*}[t]
\includegraphics[width=1.0\textwidth]{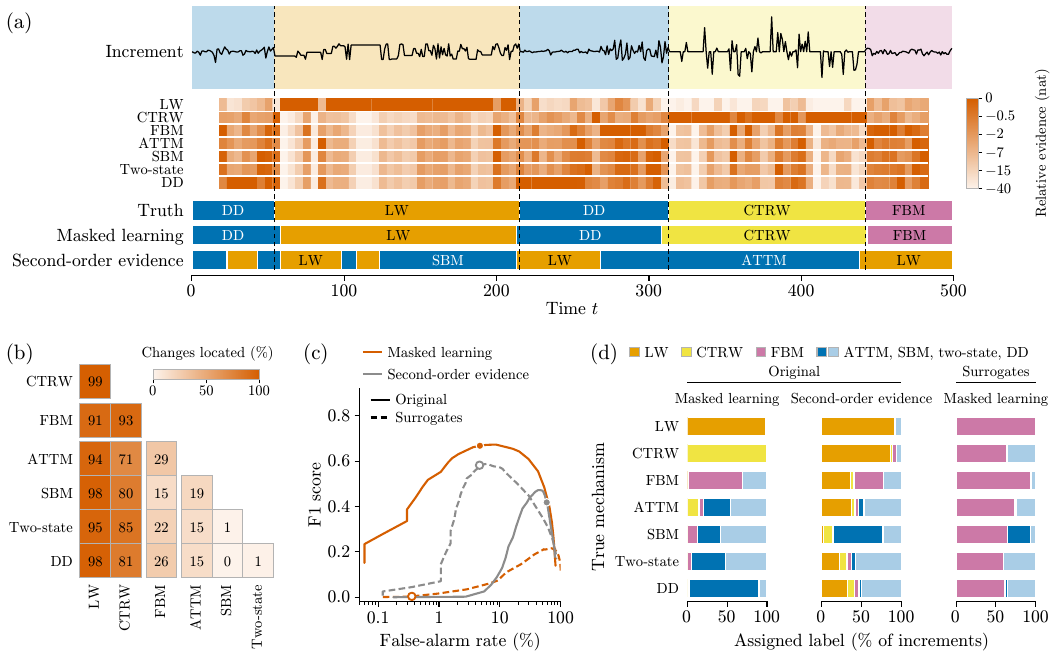}
\caption{Segmentation of multi-mechanism trajectories. The evidence of a stretch for a mechanism is the conditional log density assigned by the model of that mechanism (masked learning) or by its kriging prediction (second-order evidence); the decoder is the same. Seven mechanisms, $K=5$, $L=4$, and $\beta=0.5$ except in (c). (a) A trajectory with four mechanism changes (dashed lines). Top: increments of one coordinate on a background colored by the true mechanism. Middle: evidence from masked learning of each stretch for each mechanism, relative to the highest of the seven. Bottom: true and decoded label sequences. ATTM, SBM, two-state, and DD share one color [legend in (d)]. (b) Percentage of changes located by masked learning for each pair of adjacent mechanisms (row and column). (c) F1 score against the false-alarm rate as $\beta$ varies, on the 3360 test trajectories (solid) and on Gaussian surrogates of 1680 of them (dashed). Symbols mark $\beta=0.5$. (d) Labels assigned to the increments of each true mechanism, on the original trajectories (first two columns) and on the surrogates (third column). Of the two blues, the dark one marks the true mechanism and the light one another of ATTM, SBM, two-state, and DD.}
\label{fig:seg}
\end{figure*}

A single process often fails to describe an entire trajectory of a particle in a living cell: binding, confinement, and active transport change the mechanism of its motion~\cite{Arcizet2008,TorrenoPina2016}. These changes mark when and where the particle interacted with its surroundings, and the durations of the segments between them give the kinetics of the interactions~\cite{Persson2013,Weigel2013}. A description of the whole trajectory by a single process averages over the mechanisms and loses this information. Trajectory segmentation recovers it by locating the changes and identifying the mechanism of each segment~\cite{GroupReview2025,AnDi2021,AnDi2_2025}. Many methods evaluate a statistic or a classifier in a window that slides along the trajectory~\cite{LanoiseleeGrebenkov2017,Wagner2017}. A reliable estimate demands a long window, whereas a sharp location of the change demands a short one. Other methods train a neural network to predict the mechanism or its parameters at every point~\cite{Requena2023,GroupSeg2024}. Its training depends on trajectories labeled point by point. These are available only from simulations, with a prescribed number, spacing, and combination of changes. Two properties of the local conditional information are of advantage here. First, it is local: it refers to a hidden stretch of a few increments, and a few neighbors on either side provide most of it [Fig.~\ref{fig:theory}(b)]. The conditional distribution from which it is measured is therefore a signature of the mechanism on the scale of a few increments. This signature differs between mechanisms even where their second-order statistics coincide (Sec.~\ref{sec:identical}). Second, it requires no pointwise or changepoint labels: masked learning obtains the conditional distribution from trajectories of a single mechanism, with the hidden stretch itself as the target. The segmentation is {\it unsupervised} in this sense: no changepoint is ever shown to the models. On a trajectory with several mechanisms, the stretches of each segment follow the conditional distribution of the mechanism that generates them. A change is then located where the stretches cease to follow the learned distribution of one mechanism and begin to follow that of another. We now test how reliably mechanism changes are located in this way and how much of the result rests on structure beyond second order.

\subsection{Segmentation procedure and test trajectories}\label{sec:segprocedure}

For each of the seven mechanisms of Sec.~\ref{sec:settings}, one model with the network described there is trained by masked learning on standardized trajectories of this mechanism alone (SM, Sec.~S7~A). Neither the parameters of a segment nor the time since its process began can be known beforehand. The exponent of a mechanism is therefore drawn anew from its full range for every training trajectory. For the same reason, the network receives the neighbors of a hidden stretch without their position in the trajectory. The two-state and DD processes have no exponent and are defined by the parameters of Sec.~\ref{sec:settings}. A trajectory is scored on 93 adjoining stretches of $K=5$ increments. Each stretch $\mathbf{h}_i$ is hidden in turn, and the model of every mechanism predicts it from its neighborhood $\mathbf{n}_L$, as in the measurement of Sec.~\ref{sec:results}. Two choices differ from that measurement. First, a prediction is scored not by the energy score but by the log density that it assigns to the recorded stretch, \begin{equation} g_M(i)=\ln p^{M}_{\boldsymbol{\theta}}(\mathbf{h}_i\mid\mathbf{n}_L), \label{eq:evidence} \end{equation} which we call the evidence of the stretch for the mechanism $M$. The flow provides this density directly~\cite{Chen2018NeuralODE,Grathwohl2019FFJORD}, without the samples that the energy score requires (SM, Sec.~S7~A). The measurement could not use it, because the log density is unbounded where the conditional distribution concentrates on a single point, as in a waiting period of the CTRW or within a flight of the LW (Sec.~\ref{sec:information}). For segmentation this is an advantage: a stretch that falls on such a point is strong evidence for its mechanism (Sec.~\ref{sec:segresults}). Second, the neighborhood is kept short, $L=4$. A few neighbors provide most of the information [Fig.~\ref{fig:theory}(b)], and a longer neighborhood more often extends across a changepoint (SM, Sec.~S7~C).

A decoder without trained parameters turns the evidence of the 93 stretches into a segmentation. It selects the sequence of mechanism labels $M_1,\dots,M_{93}$ that maximizes $\beta\sum_i g_{M_i}(i)+\ln P(M_1,\dots,M_{93})$. A changepoint is placed wherever the label changes, and the prior $P$ depends only on the number of changepoints: the first one lowers $\ln P$ by 7.4~nat (SM, Sec.~S7~B). The weight $\beta$ is smaller than one because adjoining stretches are neighbors of one another, so that their densities are not independent factors of a likelihood. We use $\beta=0.5$, so that a first changepoint requires 14.8~nat of evidence. This value and $L=4$ were fixed in advance on separate trajectories (SM, Sec.~S7~B). For comparison, we use second-order evidence, decoded in the same way. Here the density of each model in Eq.~(\ref{eq:evidence}) is replaced by that of the kriging prediction of the same mechanism [Eq.~(\ref{eq:kriging})], estimated from trajectories of this mechanism. The test comprises 3360 trajectories of 499 two-dimensional increments. One half contains one to four mechanism changes per trajectory, 4200 in all; adjacent segments differ in mechanism, and every segment has at least 21 increments. The other half contains no mechanism change, so that every changepoint placed in it is a false alarm. Of these 1680 trajectories, 1400 are single realizations of one mechanism, 200 for each of the seven mechanisms. The remaining 280 join independent segments of one and the same mechanism and thus contain breaks but no change of mechanism. Every segment is cut from an independent trajectory that was standardized separately. A changepoint counts as located if a predicted changepoint lies less than 10 increments from it. We report the F1 score of the changepoints, which is the harmonic mean of precision and recall. The false-alarm rate is the fraction of the 1680 trajectories without mechanism change in which a changepoint is placed. We also report the labels that the increments of each mechanism receive.

\subsection{Segmentation results and changepoint detection}\label{sec:segresults}

To intuitively visualize the segmentation, we show an example in Fig.~\ref{fig:seg}(a). The trajectory consists of five segments: DD, LW, DD, CTRW, and FBM. With the conditional densities from masked learning, the decoder recovers the five segments. Each segment receives the label of its mechanism, and each changepoint lies within five increments of the true change. With second-order evidence, the same decoder fragments the trajectory. It cuts the first DD segment and the LW segment into pieces labeled alternately SBM and LW. It labels the CTRW segment and the second half of the preceding DD segment as ATTM, so the change between them is not located. Outside the LW segment, no part of the trajectory receives the label of its mechanism. To assess the segmentation beyond a single trajectory, we group the changes in the test trajectories by the pair of adjacent mechanisms. For each pair, Fig.~\ref{fig:seg}(b) gives the fraction of changes that masked learning locates. The changes that involve LW or CTRW, as all four changes of the example do, are located reliably, at a rate of 89\%. The changes among the other five mechanisms are located less often and are discussed below. The number of changepoints that the decoder places depends on the weight $\beta$. The solid curves in Fig.~\ref{fig:seg}(c) therefore show the F1 score against the false-alarm rate as $\beta$ varies, for both kinds of evidence. The curve of masked learning rises higher than that of second-order evidence and peaks at a much lower false-alarm rate. At $\beta=0.5$, marked by the symbols, it reaches an F1 score of 0.67 with a false-alarm rate of 4.7\%, against 0.42 with 59\% for second-order evidence. The false-alarm rate of masked learning stays low even for the two-state and DD processes, in which the diffusivity changes with time while the mechanism stays the same. Of the 400 test trajectories that are single realizations of these two processes, masked learning places a changepoint in 3, and second-order evidence in 386.

The difference between the mechanism pairs in Fig.~\ref{fig:seg}(b) originates in the evidence of the individual stretches. The middle part of Fig.~\ref{fig:seg}(a) shows this evidence for the example trajectory: the log density that each model assigns to each stretch, relative to the highest. In the LW and CTRW segments, the model of the true mechanism stands out, several nats above the others. During a flight the increments of a stretch repeat those of its neighbors, and during a waiting period they vanish, so the conditional distribution has all or nearly all of its weight on a single point (SM, Sec.~S4). A stretch within a flight puts the LW model ahead of all others by about twice the evidence that the decoder requires for a changepoint, so one or two stretches suffice. In the DD and FBM segments, by contrast, the models of FBM, ATTM, SBM, two-state, and DD typically lie within a nat or two of each other. The evidence for a change between these mechanisms must accumulate over a segment, which is often too short. This limit lies in the exact conditional densities of single stretches, not in the learned models. In a separate test with trajectories composed of Brownian motion and the two-state and DD processes, the decoder locates the changes with an F1 score of 0.27 when it uses the exact conditional densities and of 0.21 when it uses the learned ones (SM, Sec.~S7~D). Where a change is not located, the labels still identify the kind of local structure. The labels assigned to the increments of each mechanism are shown in Fig.~\ref{fig:seg}(d), in the first column for masked learning. The increments of LW and CTRW receive their own label almost without exception, and those of FBM in two thirds of the cases. Those of ATTM, SBM, two-state, and DD, in which the motion is Brownian with a time-varying diffusivity, are mostly labeled as one of these four. With these four mechanisms counted as one class, 89\% of all increments are labeled correctly, against 55\% with second-order evidence (second column). For the CTRW, 99.8\% of the increments receive their own label with masked learning, against 2.5\% with second-order evidence.

It remains to explain why masked learning produces so few false alarms where the diffusivity changes. In a two-state, DD, or SBM trajectory the diffusivity changes all along its course, and each change could be taken for a change of mechanism. What prevents this is the neighborhood of each stretch: it lets the decoder distinguish a change of state within a mechanism from a change of mechanism. To isolate what the neighbors contribute, we trained a second set of models that can also predict a stretch without neighbors. We decoded one half of the test trajectories with these models at $L=0$ and at $L=4$. Without neighbors, the decoder cuts trajectories of all three processes, and in the two-state process it does so where the diffusivity switches. With four neighbors on each side, it leaves them whole. Of 100 two-state trajectories without any change, 35 are cut at $L=0$ and none at $L=4$. Of the 59 changepoints placed at $L=0$, 57 lie within ten increments of a switch of the diffusivity. For SBM and DD, 14 and 15 of 100 trajectories are cut at $L=0$ and none at $L=4$. Over all trajectories without mechanism change, the false-alarm rate falls from 17\% to 4.0\% (SM, Sec.~S7~C). The reason is the one found in Sec.~\ref{sec:origin}: the neighbors reveal the current state of the process. Without them, a model can judge a stretch only by its amplitude, which follows the diffusivity. With them, every model takes the current diffusivity from the neighbors. What remains to be compared is the way the stretch continues its neighbors, which is set by the mechanism. The kriging prediction gains nothing from the neighbors of these processes, because their increments are uncorrelated. The information that separates the state of a process from its mechanism is therefore local conditional information beyond second-order statistics.

Finally, we test whether the advantage of masked learning over second-order evidence comes from structure beyond second-order statistics. In one half of the test trajectories, we replace every segment by a Gaussian surrogate~\cite{Theiler1992}: a stationary Gaussian sequence with the covariance of the mechanism of that segment (SM, Sec.~S7~E). The changes, the models, and the decoder stay the same. The dashed curves in Fig.~\ref{fig:seg}(c) and the third column of Fig.~\ref{fig:seg}(d) show the results for these surrogates. Masked learning treats them as what they are. It labels most increments as FBM, the only one of the seven mechanisms with stationary Gaussian increments, and places almost no changepoint: at $\beta=0.5$ its F1 score falls to 0.004. Second-order evidence, which uses only what the surrogates preserve, reaches 0.58. The advantage of masked learning thus lies in the structure that the surrogates remove.

\subsection{Segmentation of experimental trajectories}\label{sec:segexperiment}

\begin{figure*}[t]
\includegraphics[width=1.0\textwidth]{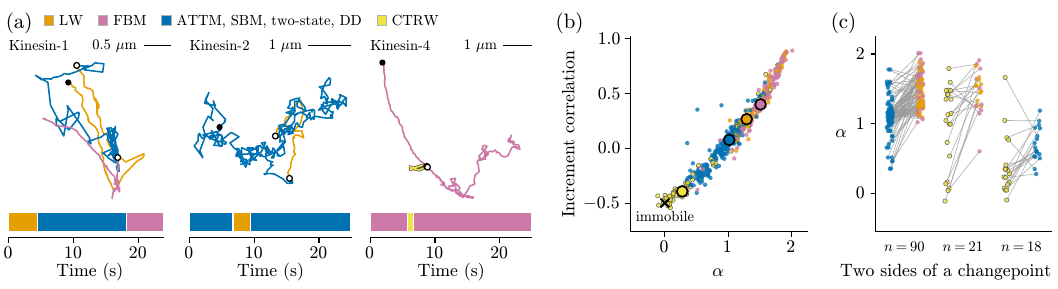}
\caption{Segmentation of experimental trajectories of quantum dots transported by kinesin motors~\cite{Katrukha2017}. The evidence comes from the seven models trained on trajectories with localization error; $K=5$, $L=4$, and $\beta=0.5$. Colors mark the decoded labels; ATTM, SBM, two-state, and DD share one color, as in Fig.~\ref{fig:seg}. (a) Three trajectories in the plane (top) and their labels along time (bottom). Filled circles mark the first position and open circles the changepoints. Each trajectory has its own scale bar. (b) MSD exponent $\alpha$, fitted to lags of one to four increments, and correlation of successive increments for every segment of at least 20 increments; a segment is a run of one color. Large circles mark the medians for each color. The cross marks an immobile particle with localization error, for which $\alpha=0$ and the correlation is $-1/2$. (c) Exponents $\alpha$ of the two segments that meet at a changepoint, joined by a line, for the changepoints with at least 20 increments on either side. From left to right: changepoints between blue and LW or FBM, between CTRW and LW or FBM, and between CTRW and blue; the segment of the color named first is on the left, and $n$ is the number of changepoints. Changepoints between LW and FBM are not shown.}
\label{fig:exp}
\end{figure*}

Unlike the test trajectories, experimental trajectories carry a localization error in every recorded position~\cite{SavinDoyle2005,Berglund2010,Michalet2010}. This error blurs the local structure that gave the strongest evidence in Sec.~\ref{sec:segresults}: the increments of a waiting period no longer vanish, and those of a flight no longer repeat. Sec.~S5 of the SM quantifies how much of the local conditional information of each process survives as the error grows. We therefore train the seven models anew on simulated trajectories with a Gaussian localization error, whose standard deviation is drawn for every trajectory between zero and its root-mean-square increment (SM, Sec.~S7~F). The second-order evidence is estimated from the same trajectories, and the decoder, $\beta$, and $L$ remain unchanged. On 1680 simulated trajectories with such errors, masked learning reaches an F1 score of 0.40 with a false-alarm rate of 14\%, against 0.28 with 28\% for second-order evidence (SM, Sec.~S7~F).

We apply these models to trajectories of quantum dots transported by kinesin motors of four families in living cells~\cite{Katrukha2017}, recorded at 50~ms per frame. Cut into pieces of at most 500 positions, 271 tracks give 488 trajectories of 100 to 499 increments (SM, Sec.~S7~G). In experiments, the motion of a particle is usually described by a few types: directed, diffusive, confined, and immobile~\cite{AnDi2_2025}. We nevertheless keep the seven mechanisms and train no model of these types, for two reasons. First, the procedure of Sec.~\ref{sec:segprocedure} is then tested as it was developed, with no model built for a particular experiment. Second, three of these types have a counterpart among the local structures of the mechanisms. Directed motion corresponds to a flight of the LW or to an FBM with $\alpha>1$, an immobile state to a waiting period of the CTRW, and diffusive motion to ATTM, SBM, two-state, and DD, which are locally Brownian. A label thus names the local structure of a part of the trajectory, not the origin of its motion in the cell. Fig.~\ref{fig:exp}(a) shows three of the trajectories in the plane and along time, colored by the decoded labels. The labels follow this correspondence. The parts in which the particle moves persistently in one direction are labeled LW or FBM, and the parts without a direction as one of ATTM, SBM, two-state, and DD. Where the particle stays in place, the label is CTRW. In total, the decoder places 189 changepoints in 105 of the 488 trajectories. Trajectories without motor-driven transport are left whole almost without exception: the decoder cuts none of 104 trajectories of quantum dots without a motor~\cite{Katrukha2017} and 3 of 300 trajectories of tracer beads in actin networks~\cite{Granik2019}.

Experimental trajectories carry no true labels, so we compare the decoded labels with two statistics that involve no model. As in Sec.~\ref{sec:segresults}, we count ATTM, SBM, two-state, and DD as one class. For every segment of at least 20 increments, Fig.~\ref{fig:exp}(b) shows the MSD exponent $\alpha$, fitted to lags of one to four increments, and the correlation of successive increments. The classes occupy different parts of this plane. The segments labeled CTRW lie near $\alpha=0$ and a correlation of $-1/2$, the values for an immobile particle with localization error; their median exponent is 0.27. The segments labeled as one of the four mechanisms with a time-varying diffusivity lie around $\alpha=1$, the value of normal diffusion (median 1.02), and those labeled LW or FBM are superdiffusive (medians 1.29 and 1.51). The correspondence seen in the three examples thus holds for the segments as a whole. Fig.~\ref{fig:exp}(c) tests the order of the classes within single trajectories. It compares the exponents of the two segments that meet at a changepoint, apart from the changepoints between LW and FBM. At 86\% of the 129 changepoints with at least 20 increments on either side, the two exponents are ordered as the medians of their classes. Trajectories of lysosomes~\cite{SittewelleRoyle2024} show the same order (SM, Sec.~S7~G).

Segmentation based on local conditional information thus carries over from simulated to experimental trajectories, without any label from the experiment and with no model built for it. Two limitations remain. Localization error costs evidence: as the error grows, the F1 score on the simulated trajectories falls and the false-alarm rate rises, and LW and FBM are no longer reliably distinguished (SM, Sec.~S7~F). The labels are also restricted to the seven mechanisms, none of which describes confined motion. Both leave room for improvement within the same framework. Each model is trained on a single process, and no changepoint is ever shown to it. A process for a type of motion of the system under study, such as diffusion with a drift or diffusion in a confined region, can therefore be added by training one more model. Masked learning also requires only trajectories, so such models could be trained on recordings in which one type of motion prevails; the localization error of the experiment would then be learned rather than assumed.

\section{Conclusion}\label{conclusion}

In this work, we have introduced the local conditional information as a general measure of the relation between the increments of a diffusion process, and measured it by masked learning. The local conditional information is the amount by which the neighborhood of a hidden stretch of increments lowers the uncertainty of this stretch, and the gain of kriging is its second-order part. Masked learning prescribes only a training task, the reconstruction of hidden stretches from their neighbors, and therefore requires nothing but trajectories of the process. Under a strictly proper score the estimate is a lower bound, so that any excess over the gain of kriging establishes information beyond second order, whatever the generating mechanism and whatever the network. The estimate reproduces the gain of kriging for the Gaussian processes and exceeds it for every non-Gaussian process studied, by 4.6\% to 11.6\% of the prior score at $L=128$. The second-order statistics of Brownian motion and of two Brownian-yet-non-Gaussian processes are identical, yet single trajectories can be assigned to the process that generated them. The accuracy of this assignment rises from 62\% to 94\% when each hidden stretch is predicted from a single neighbor on either side. Membrane receptors in living cells provide an experimental case: a model trained on their recorded trajectories alone exceeds the gain of kriging by 3.0\% of the prior score at $L=64$, where kriging itself gains 0.2\%.

For any process, the local conditional information measures how much the conditional distribution of a stretch varies from one neighborhood to another, and kriging accounts only for a linear variation of its mean. The theory of Sec.~\ref{sec:origin} reveals the origin of the remainder: a state, such as a diffusivity or a velocity, that persists in time, whether it belongs to the particle or to its surroundings. The neighbors act on a hidden stretch only by inferring this state at its two edges. The size of the information is set by the distribution of the state, and the persistence of the state determines how much of it the neighbors provide. A state that acts on the diffusivity creates no covariance between a stretch and its neighbors, however long it persists, so that the entire information lies beyond second order. This information is the part of the non-Gaussianity that persists in time.

Because a few neighbors provide most of the local conditional information, the conditional distributions learned by masked learning are signatures of a mechanism on the scale of a few increments. Compared along a trajectory, they yield an unsupervised segmentation, in which changes of mechanism are located without any changepoint label. On trajectories composed of seven mechanisms, this segmentation reaches an F1 score of 0.67 and a false-alarm rate of 4.7\%, compared with 0.42 and 59\% for second-order evidence. The advantage disappears on Gaussian surrogates that retain only the covariance. The neighbors reveal the current state of the process, so that a change of state within one mechanism is not taken for a change of mechanism. With localization error included in the training, the same procedure is used to segment experimental trajectories of quantum dots transported by kinesin motors, and the decoded labels agree with model-free statistics of the segments.

Two directions follow from this work. First, the theory of Sec.~\ref{sec:origin} assumes a Markovian state and independent noise, whereas the definition and the lower bound do not. The state may instead depend on the position of the particle, as in a quenched environment~\cite{Massignan2014,AkimotoBarkaiSaito2016}, or modulate correlated noise, as in the heterogeneous viscoelastic motion observed in living cells~\cite{Sabri2020,Janczura2021}. For such processes the measurement can precede the theory. Second, the prediction of a hidden stretch can be conditioned on other parts of a recording than its two-sided neighborhood. With the preceding increments alone, the measurement would tell how far the history of the motion determines its continuation. With the trajectory of a second particle added, it would tell what one particle reveals about another. Beyond stochastic trajectories, masked learning offers a general paradigm for discovery in physics. Wherever part of a recording can be hidden and reconstructed from the rest, the reconstruction measures what the rest reveals about it, with neither candidate mechanisms nor labels.

\begin{acknowledgments}
We thank Gongyi Wang for helpful discussions. This work is supported by the National Natural Science Foundation of China (Grant No. 12104147) and the industry-university cooperation project with DataComp AI (Project No. H202591470640).
\end{acknowledgments}

\section*{Data availability}
The code that generates the simulated trajectories, trains the models, and produces the results of this article is openly available~\cite{ourcodes}. The trained models are available from the authors upon reasonable request. The experimental trajectories were taken from published studies and are openly available~\cite{Granik2019,KatrukhaData,SittewelleRoyleData}.

\bibliography{Bibliography,SupplementalMaterial_ref}

\end{document}